   \documentclass{aa}
   \begin{document}

   \voffset 1 true cm

   \thesaurus{10     
              (08.01.1;  
               08.12.1;  
               10.05.1;  
               10.19.1)}

   \title{Consistency of the metallicity distributions of nearby 
              F, G and K dwarfs}

   \author{H. J. Rocha-Pinto \and W. J. Maciel}

   \offprints{H. J. Rocha-Pinto}

   \institute{Instituto Astron\^omico e Geof\'{\i}sico, Universidade de S\~ao 
              Paulo, Av. Miguel Stefano 4200, 04301-904 S\~ao Paulo SP, 
              Brazil\\ email: helio@iagusp.usp.br; maciel@iagusp.usp.br}
   
   \titlerunning{Consistent metallicity distributions amongst F, G and K 
    dwarfs}
   \authorrunning{H.J. Rocha-Pinto \& W.J. Maciel}

   \date{Received date; accepted date}
   
   \maketitle

   \begin{abstract}

   The consistency of the metallicity distributions of F, G and K 
   dwarfs is studied. We present a new metallicity distribution for K 
   dwarfs using metallicities determined from {\it uvby} photometry. There 
   is a remarkable paucity of metal-poor K dwarfs in analogy with the 
   G dwarf problem. We show that late-type dwarfs have consistent 
   metallicity distributions. We also propose preliminary corrections to 
   these distributions to take into account the contamination of 
   the {\it uvby} indices due to the chromospheric activity in these stars, 
   since around 30\% of the nearby late-type dwarfs have active 
   chromospheres. We consider the possibility that the metallicity
   distribution of cooler stars may be different from that of the hotter 
   stars due to (i) metal-enhanced star formation and (ii)  a 
   metallicity bias in the catalogue of nearby stars. We conclude 
   that these hypotheses are unlikely to produce important 
   differences in the metallicity distributions of late-type dwarfs.

   \end{abstract}

\section{Introduction}

   Thirty six years after its discovery by van den Bergh (\cite{vanden}), 
   the G dwarf problem still presents challenges to the astrophysicists 
   studying Galactic Evolution. Although several mechanisms for decreasing 
   the number of metal-poor dwarfs in the Galaxy have already been devised, 
   the shape of the metallicity distribution is generally  not very well 
   reproduced by the majority of models in the literature. In fact, given 
   the uncertainties in the data, obtaining a good fit to the G dwarf 
   metallicity distribution was less significant than to search for an 
   explanation for the paucity of metal-poor objects. However, after the 
   recent derivation of a new G dwarf metallicity distribution 
   (Rocha-Pinto \& Maciel \cite{RPM96}, hereafter RPM), the G dwarf problem 
   cannot be regarded as just the paucity of metal-poor stars, compared with 
   Simple Model predictions. RPM showed that, besides the small number of 
   metal-poor objects, there is also a small number of metal-rich dwarfs and 
   an excessive number of dwarfs with intermediate metallicities.

   These results were already predicted by Malinie et al. (\cite{malinie})
   on the basis of an inhomogeneous chemical evolution model. Infall models 
   also seem very suitable to reproduce the shape of the new metallicity 
   distribution, as shown by RPM and Chiappini et al. (\cite{cris}).

   Recently, Favata et al. (\cite{favata1}) have obtained spectroscopic 
   metallicities for a sample of 91 nearby G and K dwarfs. They found a very 
   narrow K dwarf metallicity distribution, in which no stars have [Fe/H] 
   $< -0.4$, in contrast with the broader G dwarf metallicity distribution 
   they have also derived.  They have offered two possible explanations for 
   this discrepancy: the Second Catalogue of Nearby Stars (Gliese 
   \cite{gliese}; Gliese \&  Jahrei\ss\ \cite{glijar1}; hereafter CNS2) 
   from which they have selected their sample could 
   have a metallicity bias, in the sense of favouring metal-rich stars; 
   alternatively, less  massive stars should preferably form in metal-rich 
   regions. 

   In this paper, we make an effort to derive the metallicity distribution 
   of K dwarfs in the solar neighbourhood, along the same lines followed for 
   the G dwarfs (RPM). Our main purpose is to see whether or not these 
   distributions are different from each other. As it will become clear 
   in the following sections, the metallicity distribution by RPM can be 
   taken as representative of the true late-type star metallicity 
   distribution in the solar neighbourhood. We also present preliminary  
   corrections to photometrically derived metallicity distributions 
   that take into account the effect of the chromospheric activity on 
   the {\it uvby} indices (Giampapa et al. \cite{giampapa}; Basri et al. 
   \cite{basri}; Gim\'enez et al. \cite{gimenez}; Morale et al. 
   \cite{morale}; Rocha-Pinto \& Maciel \cite{RPM98}). The contamination 
   of the photometric indices by the chromospheric activity is one of the 
   most important sources of systematic errors in photometric [Fe/H] surveys, 
   and has been often ignored.

   This paper is organized as follows: in Sect. 2, we present the selection 
   criteria for the sample of K dwarfs, and derive the corresponding 
   metallicity distribution. In Sect. 3, the derived distribution is compared 
   with the G dwarf metallicity distribution, and the consistency of the 
   metallicity distributions of late-type dwarfs of types F, G, and K is 
   considered. In Sect. 4, we present the proposed corrections owing to the 
   chromospheric activity, and apply them to both G and K dwarf metallicity 
   distributions. A discussion of the results by Favata et 
   al. (1997), especially regarding  the differences between their derived 
   distributions is given in Sect. 5.
   
\section{The K dwarf metallicity distribution}

   We have selected a preliminary sample from the Third Catalogue of Nearby 
   Stars (Gliese \& Jahrei\ss\ \cite{glijar2}; hereafter CNS3). This  sample 
   comprises around 870 objects classified as K stars. We searched for
   {\it uvby} indices for these stars in the surveys of Olsen (\cite{olsen93}, 
   \cite{olsen94}) and in the compilation by Hauck \& Mermilliod (\cite{HM90}), 
   favouring the data by Olsen when a star had measurements in  both sources. 
   Disregarding unresolved binaries, stars with variable indices, giants and 
   subgiants, our sample has been reduced to 242 objects. For some of these, 
   the spectral types available in the literature do not allow the 
   identification of the star luminosity class. In these cases, the 
   identification was made by checking the star's position on the 
   $(b-y)\times c_1$ diagram. Seventeen objects occupy a region in this 
   diagram which is mainly populated by subgiants, according to Olsen 
   (\cite{olsen84}), and were eliminated from the sample. One star 
   (\object{BD +00 3077}) was also removed from the sample, as it has a 
   colour $(b-y)=0.972$ of an M dwarf, although being classified 
   as K7 V in the CNS3.

   Metallicities were found from the calibrations of Schuster \& Nissen 
   (\cite{schuniss}) for stars bluer than $(b-y)=0.550$, and from the 
   calibration for K2--M2 dwarfs by Olsen (\cite{olsen84}) for the redder 
   stars. The calibrations by Schuster \& Nissen are assumed  to be valid 
   for $(b-y) < 0.590$. However, we decided to apply them for 
   $(b-y)<0.550$ only, since beyond this value the calibrations 
   yield spuriosly high metallicites of 0.45--0.75 dex. On the 
   other hand, the calibration by Olsen (\cite{olsen84}) is valid for 
   the range $(b-y)> 0.514$, but it is rather uncertain for 
   $(b-y)>0.550$, as it is based on a small number of stars
   with spectroscopic [Fe/H] determinations. Therefore, 
   the accuracy of the metallicity determinations 
   for the cooler stars is poorer than for the hotter objects.

   Figure \ref{ironcomp} shows the comparison between our derived 
   photometric metallicities and spectroscopic metallicities taken from 
   the literature (Cayrel de Strobel et al. 1997, Favata et al. 1997) for 
   42 dwarfs. It can be seen that the photometric and spectroscopic data 
   are in good agreement with each other, especially when data by Favata 
   et al. (1997) is used. 

      \begin{figure}
      \vspace{6cm}
      \caption[]{Comparison between the photometric and spectroscopic 
               metallicities for 42 K dwarfs. The spectroscopic data are 
               from Favata et al. (\cite {favata1}) and  Cayrel de Strobel 
               (\cite{cayrel}).}
   \label{ironcomp}
   \end{figure}

   Characterization of the disk population has been made by applying the 
   chemical criterion (see RPM for details), according to which stars with 
   ${\rm [Fe/H]} < -1.2$ are considered as halo members. From the 
   application of this criterion, 6 stars were removed from the sample, 
   which comprises 218 K dwarfs in its final form. Detailed
   data on these stars can be supplied by request to the authors. As 
   discussed by RPM, the chemical criterion is a very simplistic one and 
   does not take into account the recent results on the chemical and 
   kinematical properties of the halo and thick disk 
   (Beers \& Sommer-Larsen \cite{beers}; Gratton et al. \cite{gratton}). 
   In fact, the chemical 
   criterion is presently more traditional than astrophysical, as it 
   allows a straight comparison between our distribution 
   and previous studies in the literature. More rigorously, the 
   characterization of a pure thin disk late-type dwarf sample should be 
   made  by considering both the chemical composition and spatial 
   velocities of the stars. At present this is not possible, as
   radial velocities are available only for a few late-type 
   disk stars. 

  \begin{table}
      \caption[]{Metallicity distribution of 218 nearby K dwarfs}
         \label{kdist}
         \begin{flushleft}
{\halign{%
\hfil#\hfil&\qquad\hfil#\cr
\noalign{\hrule\medskip}
[Fe/H] & number \cr
\noalign{\medskip\hrule\medskip}
            $-1.15$ & 0    \cr
            $-1.05$ & 0    \cr
            $-0.95$ & 0    \cr
            $-0.85$ & 0    \cr
            $-0.75$ & 3    \cr
            $-0.65$ & 2    \cr
            $-0.55$ & 5    \cr
            $-0.45$ & 11   \cr
            $-0.35$ & 18   \cr
            $-0.25$ & 39   \cr
            $-0.15$ & 39   \cr
            $-0.05$ & 36   \cr
             $0.05$ & 30   \cr
             $0.15$ & 28   \cr
             $0.25$ & 6    \cr
             $0.35$ & 1    \cr
\noalign{\medskip\hrule}}}
         \end{flushleft}
   \end{table}

   Particular care must be taken in the sense of avoiding any bias towards 
   metal-poor stars in our sample. Some bias could be produced by 
   intrinsic biases in the {\it uvby} databases we have used. From the 
   218 K dwarfs in our final sample, 138 have photometric data from 
   Olsen (\cite{olsen93}), 40 from Olsen (\cite{olsen94}) and 
   40 from Hauck \& Mermilliod (\cite{HM90}). It is difficult to 
   investigate the presence of any bias in the compilation by 
   Hauck \& Mermilliod, as it contains objects from several 
   heterogeneous sources. On the other hand, the samples in Olsen's 
   papers are very well described and different subsamples  
   are easily identified, particularly in Olsen (\cite{olsen93}).
   Three subsamples of this last catalogue are present in our 
   sample: G5-type HD stars, calibration stars and high-velocity 
   stars. Biases could be present in the calibration stars 
   due to selection effects, and high-velocity stars which 
   are likely to be old metal-poor stars. 
   Of the 138 stars in our sample taken from Olsen (\cite{olsen93}), 
   38 are G5-type HD stars, 77 are calibration stars and 
   23 are high-velocity stars. The average metallicity of G5-type stars  
   is around $-0.19$ dex, while the calibration and high-velocity 
   stars have average metallicites of $-0.10$ and $-0.07$ dex, 
   respectively. The average metallicity of 
   the stars coming from the catalogues 
   of Olsen (\cite{olsen94}) and Hauck \& Mermilliod (\cite{HM90}) 
   is around $-0.15$ dex. The standard deviation of the metallicity 
   distributions of all these subsamples is 0.21--0.23 dex. Therefore,
   no bias towards metal-poor objects is likely to be present
   in our sample. The differences in the metallicity 
   distribution of the subsamples may suggest a small bias towards 
   metal-rich objects. However, these differences may be caused by 
   the fact that the subsamples have different $(b-y)$ ranges, 
   some of which depend more strongly on the different metallicity 
   calibrations we used.

   The resulting metallicity distribution is presented in Table \ref{kdist}. 
   It can be seen that no stars have ${\rm [Fe/H]} < -0.80$, in excellent 
   agreement with the previous results by RPM. The data in Table 
   \ref{kdist} show that the `G dwarf problem' is not a characteristic 
   of the G dwarfs only, ruling out all previous arguments that the paucity 
   of metal-poor dwarfs could be caused by the non-legitimacy 
   of the G dwarfs as representative of the long-lived stars (see 
   Rocha-Pinto \& Maciel \cite{RPM97a}). In fact, the existence of a 
   {\it K dwarf problem} confirms that the paucity of metal-poor long-lived 
   stars is a real feature of the galactic disk. It is interesting to note 
   that, according to Worthey et al. (\cite{worthey}), the G 
   dwarf problem could even be an universal consequence of the 
   evolution of galaxies.

   \section{Comparison of the metallicity distributions of F, G, and K dwarfs}

   Figure \ref{GKcomp} shows a comparison between our K dwarf metallicity 
   distribution and that of the G dwarfs (RPM). It can be seen that there is a
   very good agreement between these distributions, with only some small differences 
   in the range $-0.7 < {\rm [Fe/H]} < -0.4$, and in the amplitude of the 
   peak around ${\rm [Fe/H]} \approx -0.25$. Therefore, there seems to be no essential
   difference in the distributions of hotter and cooler dwarfs, in opposition to 
   the findings by Favata et al. (\cite{favata1}). This conclusion is supported by
   many independent metallicity distributions in the literature, which  agree with 
   the G dwarf metallicity distribution by RPM. This is shown in Figure \ref{othercomp}, 
   where we show, besides the metallicity distribution of RPM:
   \begin{enumerate}
     \item The metallicity distribution of the F dwarf sample studied by Twarog 
        (\cite{twarog}), comprising 936 stars, after applying corrections 
        due to stellar evolution and scale height, assuming the Salpeter initial mass 
        function (IMF). Twarog's (\cite{twarog}) sample was built with the primary 
        purpose of studying the age--metallicity relation. It is composed exclusively 
        by F dwarfs, selected by $T_{\rm eff}$ range, and is expected to be 
        representative of our vicinity. Metallicities are found from {\it uvby} 
        photometry, but using a very simple calibration in which [Fe/H] 
        depends linearly on $\delta m_1$.

     \item The metallicity distribution of Wyse \& Gilmore (\cite{wyse}), with 128 
         F and G dwarfs. Wyse \& Gilmore (\cite{wyse}) use the same photometric 
         calibrations as RPM. The major difference between these works is that 
         Wyse \& Gilmore (\cite{wyse}) have used photometric data by Olsen (\cite{olsen83}),
         while RPM have used the more  recent data from Olsen (\cite{olsen93}). This 
         last paper is specifically concerned with G stars, while Olsen (\cite{olsen83}) 
         gives more attention to stars ranging from A0 to G0. Therefore, their 
         metallicity distribution includes some late F dwarfs, apart from the G dwarfs.

     \item The metallicity distribution of Flynn \& Morell (\cite{flynn}), 
         comprising 179 G and K dwarfs, after applying the chemical criterion.
         They have built their sample from G and K dwarfs, listed in  CNS3, with 
         $(R-I)$ measurements and Geneva photometric indices available in the literature. 
         Their sample has 179 stars with ${\rm [Fe/H]} \ge -1.2$ after applying the 
         chemical criterion, from which 97 are G dwarfs and 82 are K dwarfs. 
         In order to improve the statistics of their database, we have used the metallicity 
         distribution for their combined sample of G and K dwarfs. 

     \item The metallicity distribution derived by Rocha-Pinto \& Maciel 
         (\cite{RPM98}), based on the chromospheric activity survey 
          (Soderblom \cite{soder}; Henry et al. \cite{HSDB}),  with 730 dwarfs 
          of types late F, G and early K. All stars in the chromospheric activity 
          survey are expected to be located within 50 pc from the Sun, and are mostly 
          G dwarfs, with some late F and early K dwarfs. Str\"omgren photometric indices 
          for these stars were taken from Olsen (\cite{olsen83}, \cite{olsen93},  
          \cite{olsen94}) and used to find 
          metallicities adopting the same calibrations used here.

    \end{enumerate}

    \begin{figure}
      \vspace{6cm}
      \caption[]{Comparison between the metallicity distributions for K dwarfs 
               (this work) and G dwarfs (RPM).}
         \label{GKcomp}
   \end{figure}

   \begin{figure}
      \vspace{6cm}
      \caption[]{Comparison of the G dwarf metallicity distribution (RPM) and other 
                distributions in the literature.}
         \label{othercomp}
   \end{figure}

   All these distributions use metallicities estimated by photometric data. However, 
   they differ in the selection criteria and calibrations used. In spite of these 
   differences, the agreement of the metallicity distributions (Figures 2 and 3) is 
   very good. The fraction of stars with [Fe/H] $<-0.40$, for each distribution, is 
   presented in Table \ref{fraction}. From these data, it can be estimated that around 
   (22 $\pm$ 7)\% of the late-type dwarfs in our neighbourhood should have 
   metallicities lower than $-0.40$ dex

     \begin{table}
      \caption[]{Fraction of dwarfs with [Fe/H] $< -0.40$ in the metallicity 
           distribution}
         \label{fraction}
         \begin{flushleft}
{\halign{%
#\hfil & \qquad\hfil$#$\hfil \cr
\noalign{\hrule\medskip}
 This work  & 9.6\% \cr
 Twarog (\cite{twarog}) & 22\% \cr
 Wyse \& Gilmore (\cite{wyse}) & 20.3\% \cr
 RPM & 18.4\% \cr
 Flynn \& Morell (\cite{flynn}) & 31.3\% \cr
 Rocha-Pinto \& Maciel (\cite{RPM98}) & 13.2\% \cr
\noalign{\medskip\hrule}}}
         \end{flushleft}
   \end{table}

   Note also that all distributions, except that by Flynn \& Morell (1997), 
   show a prominent single peak around $-0.20$ dex. As shown by Rocha-Pinto \& Maciel 
   (\cite{RPM97b}), this feature could be explained by an intense star formation era 
   from 5 to 8 Gyr ago. Therefore, the main conclusion that can be drawn from the 
   comparisons above is that there is a remarkable consistency amongst the distributions 
   of F, G and K dwarfs. This consistency could only be attained if the chemical enrichment 
   and star formation history  \emph{have been essentially the same for all 
   late-type dwarfs}.

\section{Correction factors owing to chromospheric activity}

   The raw data of the metallicity distributions are often subject to a variety 
   of corrections due to observational errors, cosmic scatter and scale height effects. 
   When a sample has stars with lifetimes lower than the disk age, corrections due to 
   stellar evolution must also be applied. Such corrections are needed to convert the 
   \emph{observed} metallicity distribution into the \emph{true} distribution. 

   For a distribution based on spectroscopic [Fe/H], these  corrections 
   are generally sufficient. However, for photometric distributions 
   there is an additional correction which has been totally neglected in 
   past studies. This correction is needed in order to take into account the effects of 
   the chromospheric activity on the photometric indices.
 
   By studying the metallicity distribution in a sample of 730 late-type dwarfs with 
   varying levels of chromospheric activity, Rocha-Pinto \& Maciel (\cite{RPM98}) 
   have shown that, for the active stars, the difference between the spectroscopic and the 
   photometric metallicity increases systematically as a function of the stellar 
   activity. This result is a consequence of the $m_1$ deficiency, which is more 
   pronounced in active binaries (Gim\'enez et al \cite{gimenez}), but actually seems 
   also to be present in normal active stars (Giampapa et al. \cite{giampapa}; Basri 
   et al. \cite{basri}; Morale et al. \cite{morale}). A metallicity distribution 
   that does not take into account this effect will be biased towards metal-poor 
   stars. The elimination of identified active stars from the sample is not an ideal 
   solution to this problem as, in single late-type dwarfs, the activity is linked 
   to the stellar age (Soderblom et al. \cite{soder91}). Samples free of active stars 
   will be also free of young stars, which will introduce another bias, in the sense 
   of avoiding the expected metal-richer dwarfs. Even if there was no  relation between 
   age and activity, there would always remain some unidentified active stars in 
   the photometric surveys, as we do not know how to identify such stars from their 
   indices. The only way to keep a minimum compromise between the achievement of 
   a non-biased sample and an accurate metallicity distribution is to make use of approximate 
   corrections for the effects of the chromospheric activity.

   The corrections we are proposing assume that all active stars, for which
   the chromospheric index  $\log R'_{\rm HK}>  -4.75$ (Soderblom et al. 1991),
   have photometric metallicities lower than the spectroscopic values by a constant 
   amount $\Delta$.  In fact, $\Delta$ is likely to depend on $\log R'_{\rm HK}$, but 
   for the sake of simplicity we shall adopt here an average value given by
     
   \begin{equation}
   \bar\Delta={\int^\infty_{-4.75}\chi(\log R'_{\rm HK})\Delta(\log R'_{\rm HK})\,{\rm d}
   \log R'_{\rm HK}\over \int^\infty_{-4.75}\chi(\log R'_{\rm HK})\,{\rm d}\log R'_{\rm HK}},
   \label{deltamean}
   \end{equation}
   
   \noindent
   where $\chi(\log R'_{\rm HK})$ is the distribution of stellar chromospheric activity, 
   that can be found from the combined data of Soderblom (\cite{soder}) and Henry et al. 
   (\cite{HSDB}), and $\Delta$ is estimated by using Eq. (5) of Rocha-Pinto \& Maciel 
   (\cite{RPM98}). Using Eq. (\ref{deltamean}), we have $\bar\Delta=0.149$ dex.

   The normalized photometric metallicity distribution of the active stars, 
   ${\cal D}({\rm [Fe/H]})$, from Rocha-Pinto \& Maciel (\cite{RPM98}), is shown in 
   Table \ref{xdist}. Instead of identifying the active stars in the data sample, 
   the approach we have taken here assumes that a fraction $c$ of the total number of 
   stars in the sample ($N_{\rm tot}$) are active stars. Therefore, the number of active 
   stars in each metallicity bin is $cN_{\rm tot}{\cal D}({\rm [Fe/H]})$, and to correct 
   the metallicity distribution, these active stars should be allocated to more metal-rich 
   bins by an amount of $\bar\Delta$.

     \begin{table}
      \caption[]{Metallicity distribution for active stars and corrections}
         \label{xdist}
         \begin{flushleft}
{\halign{%
\hfil#\hfil & \quad\hfil#\hfil & \quad\hfil#\hfil & \quad\hfil#\hfil & \quad\hfil#\hfil \cr
\noalign{\hrule\medskip}
[Fe/H] & ${\cal D}{\rm [Fe/H]}$ & $r$ & $r_{\rm G}$ & $r_{\rm K}$ \cr
\noalign{\medskip\hrule\medskip}
            $-1.15$ & 0  & 0 & 0 & 0 \cr
            $-1.05$ & 0  & 0 & 0 & 0 \cr
            $-0.95$ & 0  & 0 & 0 & 0 \cr
            $-0.85$ & 0  & 0 & 0 & 0 \cr
            $-0.75$ & 0.00508 & $-0.00009$ & $-0.01$ & $-0.01$  \cr
            $-0.65$ & 0.01015 & $-0.00096$ & $-0.08$ & $-0.06$ \cr
            $-0.55$ & 0.01523 & $-0.00685$ & $-0.58$ & $-0.44$  \cr
            $-0.45$ & 0.02030 & $-0.03097$ & $-2.63$ & $-2.00$ \cr
            $-0.35$ & 0.09645 & $-0.08674$ & $-7.37$ & $-5.60$ \cr
            $-0.25$ & 0.20305 & $-0.14168$ & $-12.04$ & $-9.14$ \cr
            $-0.15$ & 0.26396 & $-0.10795$ & $-9.17$ & $-6.97$ \cr
            $-0.05$ & 0.20305 & 0.02736 & 2.32 & 1.77 \cr
             $0.05$ & 0.09645 & 0.13495 & 11.46 & 8.71 \cr
             $0.15$ & 0.07107 & 0.12700 & 10.79 & 8.19 \cr
             $0.25$ & 0.01015  & 0.06311 & 5.36 & 4.07 \cr
             $0.35$ & 0.00508  & 0.01884 & 1.60 & 1.22 \cr
             $0.45$ & 0 & 0.00352 & 0.30 & 0.23 \cr
\noalign{\medskip\hrule}}}
         \end{flushleft}
   \end{table}

  The fraction $c$ is likely to depend on the spectral type considered, as 
  the chromospheric activity is thought to be caused by the interaction 
  between the stellar rotation and the convection in the stellar envelope. The decrease 
  of the outer convective zone towards hotter stars indicates that young hotter 
  stars do not show much activity (Elgar\o y et al. \cite{elgaroy}). 
  For a sample centered on G dwarfs, we can take $c=0.296$ 
  as a good value, according to Henry et al. (\cite{HSDB}). 
 
  Table \ref{xdist} also presents the normalized corrections $r$ to 
  the metallicity distribution. The numbers in the table were found 
  by the subtraction of ${\cal D}{\rm [Fe/H]}$ from a gaussian curve 
  fitted to this distribution with a mean shifted by $\bar\Delta$. 
  These corrections are to be 
  multiplied first by $cN_{\rm tot}$, before they can be added to 
  the metallicity distribution, and \emph{before} the application 
  of any other corrections due to observational errors, cosmic scatter, 
  stellar evolution or scale height. 
  
  The absolute corrections to the G dwarf metallicity distribution 
  of RPM and the K dwarf distribution derived in this work are shown 
  in the last columns of Table \ref{xdist}, where 
  $r_K = r c N_{\rm tot}(K)$ and $r_G = r c N_{\rm tot}(G)$ with
  $N_{\rm tot}(K) = 218$ and $N_{\rm tot}(G) = 287$. Note that we 
  have assumed the same values for $c$ and $\bar\Delta$ for G and K 
  dwarfs, as there is no information about their dependence on the 
  stellar mass.

  It should be stressed that these corrections are valid 
  only for distributions binned by 0.1 dex, with  each bin centered at the metallicities 
  presented in the first column of Table \ref{xdist},  and for [Fe/H] determined 
  by Str\"omgren photometry. In order to apply them to a distribution binned in a 
  different way, we provide the equations below: 

  \begin{equation}
  r_X = 0.296\,\delta z\, N_{\rm tot}(X)\left[G({\rm [Fe/H]}-\bar\Delta)-G({\rm [Fe/H]})\right],
  \label{r}
  \end{equation}

  \noindent
  where $\delta z$ is the bin size in dex, assumed constant, and

  \begin{equation}
  G({\rm [Fe/H]}) = {1\over\sigma\sqrt{2\pi}} \exp\left[-{\left({\rm [Fe/H]}-\mu\right)^2\over 
  2\sigma^2}\right]
  \label{gauss}
  \end{equation}

  \noindent
  is the gaussian fit to the normalized distribution in Table \ref{xdist}. According 
  to this fit, $\mu=-0.143$ and $\sigma=0.152$.

  For distributions based on different photometric systems, a new value for 
  $\bar\Delta$ should be computed, since the extent of chromospheric activity effects 
  on the photometric indices depends on the spectral range sampled by the filters, 
  as well as on their transmission functions. This could  be an explanation for the fact 
  that the metallicity distribution of Flynn \& Morell is somewhat different 
  from the others (see Figure \ref{othercomp}), as this distribution uses Geneva 
  photometry, and the indices of the calibrations can be affected in a different way 
  from the {\it uvby} indices,  which are used by all other distributions in 
  Figure \ref{othercomp}.

   \begin{figure}
      \vspace{5cm}
      \caption[]{$\delta m_1$ as a function of the chromospheric activity for 
               the active stars in the sample of Rocha-Pinto \& Maciel (\cite{RPM98}). 
               G dwarfs and K dwarfs are marked by solid and open circles, respectively.}
         \label{m1active}
   \end{figure}

  Morale et al. (\cite{morale}) report that, in active K dwarfs, $\delta m_1$ is 
  systematically greater than in active G dwarfs as a function of the stellar activity, 
  which would indicate a greater $\bar\Delta$ for those stars. This is not confirmed for 
  the active stars in the sample studied by Rocha-Pinto \& Maciel (\cite{RPM98}), as  can 
  be seen from Fig. \ref{m1active}. This plot shows that the $m_1$ deficiency, reflected 
  in a larger value for $\delta m_1$, is about the same for G and K dwarfs, as a function 
  of the activity. However, the stars analyzed by Morale et al. (1996) are generally much 
  more active than ours, as they were detected by the X-ray flux-limited {\it Einstein} 
  Extended Medium Sensitivity Survey (Gioia et al. \cite{gioia}). 

     \begin{table}
      \caption[]{Activity indices for common stars in the Einstein and 
               chromospheric activity survey}
         \label{einstein}
         \begin{flushleft}
{\halign{%
\hfil# & \quad\hfil$#$\hfil & \quad\hfil$#$\hfil  \cr
\noalign{\hrule\medskip}
Name & \log R'_{\rm HK} & \log(f_{\rm X}/f_{\rm V})\cr
\noalign{\medskip\hrule\medskip}
            \object{HD 105} & -4.36  & -3.58 \cr
            \object{HD 166} & -4.33  &  -3.43 \cr
            \object{HD 25680} & -4.54  &  -3.93 \cr
            \object{HD 97334} & -4.40  & -4.06 \cr
  \noalign{\medskip\hrule}}}
         \end{flushleft}
   \end{table}

  This can be verified from the data in Table \ref{einstein} where we compare the 
  activity indices, $\log R'_{\rm HK}$ and $\log(f_{\rm X}/f_{\rm V})$, in the 
  chromospheric activity and  {\it Einstein} surveys, respectively, for the four 
  stars in common to these surveys. The bulk of the active stars, according to the 
  distribution function $\chi(\log R'_{\rm HK})$, has $\left\langle\log R'_{\rm HK}\right
  \rangle\approx -4.50$, which from  the values in Table \ref{einstein} would correspond to 
  $\log(f_{\rm X}/f_{\rm V})\approx -3.9$ or lower. Thus, our Fig. \ref{m1active} does not 
  rule out the conclusions by Morale et al. (\cite{morale}). Note that our most active 
  stars, that would have $\log(f_{\rm X}/f_{\rm V})\approx 
  -2.8$ if we extrapolate the relation for the stars from Table \ref{einstein}, have 
  $\delta m_1\approx 0.07$ in good agreement with Figure 3 by Morale et al.. We can 
  see that at $\log(f_{\rm X}/f_{\rm V})\approx -3.0$, the G and K dwarfs still present 
  similar $\delta m_1$ indices.

  From the considerations above, we can conclude that, only for the most active dwarfs, 
  the cooler stars will present larger $\Delta$ compared to the G dwarfs. From the 
  function $\chi(\log R'_{\rm HK})$, these very active stars comprise around 
  5\% of the active stars we are dealing with (that is, $0.05cN_{\rm tot}$ stars), 
  so that their influence on the metallicity distribution will  be negligible, and 
  our hypothesis for equal $c$ and $\bar\Delta$ is fairly reasonable.

\section{Metal-enhanced star formation of K dwarfs or a biased catalogue?}
                                                
   \begin{figure}
      \vspace{9.5cm}
      \caption[]{Comparison between the metallicity distributions for K dwarfs 
               (this work) and G dwarfs (RPM), and Favata et al.'s distributions.
              }
         \label{favcomp}
   \end{figure}

  In the last few years, several works have investigated the observational aspects of the 
  G dwarf problem (Wyse \& Gilmore \cite{wyse}; Rocha-Pinto \& Maciel \cite{RPM96}, 
  \cite{RPM97a}; Flynn \& Morell \cite{flynn}). All these works have followed the steps 
  delineated by Pagel \& Patchett (\cite{pagel}) for the selection of a unbiased 
  metallicity distribution of long-lived dwarfs, by choosing stars in a volume limited 
  sample and using photometric metallicities. 

  The recent paper by Favata et al. (\cite{favata1}) also analyzes the metallicity 
  distribution of the solar neighbourhood. The major novelty of this work is that the 
  authors made the first attempt to systematically study the local metallicity 
  distribution by using spectroscopic metallicities. In fact, the first local 
  spectroscopic metallicity distribution was made by Rana \& Basu (\cite{ranabasu}). 
  However, their selection criteria were not approppriate to define a unbiased sample, 
  and their metallicity database was largely heterogeneous. Recently, some papers have 
  also made use of a spectroscopic metallicity distribution from the data of Edvardsson 
  et al. (\cite{Edv}). However, this distribution cannot be taken as representative either, 
  as Edvardsson et al. have selected their stars in order to have nearly equal numbers of 
  them in pre-determined metallicity bins. 

  The results by Favata et al. (\cite{favata1}) are quite peculiar: stars hotter than 
  5100 K present metallicities spanning the whole range of [Fe/H] values expected for the 
  disk, whereas amongst the cooler objects, no stars show  ${\rm [Fe/H]} < -0.40$ dex. Their 
  sample comprises 91 stars, 65 of which are considered as G dwarfs and 26 are K dwarfs, 
  their separation being made at 5100 K.

  The authors present two alternative hypotheses to explain the lack of cool 
  metal-poor stars:
   \begin{enumerate}
     \item Low mass stars would preferably form in higher metallicity clouds, due to the 
         efficient cooling driven by the radiation of molecules containing metals.
     \item The Catalogue of Nearby Stars could have a metallicity bias, in the sense of 
         favouring metal-rich stars amongst the cooler ones.
    \end{enumerate}

  In what follows, we shall examine these hypotheses separately.

  \subsection{Metal-enhanced star formation of K dwarfs}

  The first hypothesis resembles the metal-enhanced star formation model (MESF;
  Talbot \& Arnett \cite{talbot73}; Talbot \cite{talbot74}; see also Tinsley \cite{tinsley75}, 
  \cite{tinsley80}). This model was proposed to explain the lack of metal-poor G 
  dwarfs, when the G dwarf problem was identified. The idea of Favata et al. (\cite{favata1}), 
  although not explicitly stated in this way, is that stars of progressively lower masses are 
  generally born with metallicities above than average, just like a mass-dependent 
  metal-enhanced star formation.

  There are problems with this hypothesis. If MESF could produce a lack of metal-poor 
  K dwarfs compared to G dwarfs, then the same reasoning indicates that there would be 
  a paucity of metal-poor G dwarfs compared to F dwarfs, and so on. It is not 
  possible to test this hypothesis using stars earlier than F0, since the older earlier 
  stars have already evolved away from the main sequence. However, the F dwarf 
  metallicity distribution corrected by stellar evolution (Twarog \cite{twarog}) is not 
  different from the distribution of the G dwarfs in the metal-poor range (see Figure 
  \ref{othercomp}). The F dwarf metallicity distribution could have another intrinsic bias 
  towards metal-rich stars due to the accretion of Jupiter-mass planets (Laughlin \& Adams 
  \cite{laugh}). However, the extent of these effects is not presently known. Moreover, 
  as there is a metallicity gradient in the Galaxy (see for example Maciel and K\"oppen
  \cite{maciel}), the fraction of cooler dwarfs  related to the other stars should increase 
  towards the Galactic center. Studies of the variation of the IMF as a function of 
  galactocentric radius show just the opposite (Scalo \cite{scalo}; Matteucci \& 
  Brocato \cite{matbro}).

  Figure \ref{favcomp} compares the metallicity distributions found by Favata et al. 
  (\cite{favata1}) with the G dwarf (RPM) and our present K dwarf metallicity distributions, 
  after the application of the corrections due to chromospheric activity. These corrections 
  were not applied to these distributions in the previous figures, since we were 
  comparing photometric distributions, which are expected to be affected in the same way by 
  chromospheric activity. However, to compare a photometric distribution with a spectroscopic 
  one, the corrections in Table \ref{xdist} are needed. The G dwarf metallicity distributions 
  show a good agreement (upper panel of Figure \ref{favcomp}), except for [Fe/H] $>+0.10$, 
  where the distribution by Favata et al. (\cite{favata1})  shows a larger number of 
  metal-rich stars. The same occurs in the K dwarf distribution
  (lower panel of Figure \ref{favcomp}). Note also the lack of 
  metal-poor K dwarfs in the sample by Favata et al. (\cite{favata1}) compared to ours. 
  This difference is not likely to be caused by errors in the photometric calibrations 
  we have used, since Figure \ref{ironcomp} demonstrates the good agreement with the 
  spectroscopic metallicities, which is even closer for their data.

  The MESF model was not successful in giving a reasonable explanation to the G dwarf 
  problem,  as it requires both very large chemical inhomogeneities in the interstellar 
  medium and very inefficient star formation in metal-poor regions (Tinsley \cite{tinsley80}).
  Our present knowledge of star formation and initial mass function corroborates this, 
  as we shall show below.

  Padoan et al. (\cite{padoan}) have recently presented analytical expressions for the 
  initial mass function (IMF) taking into account the dependence of the star formation on 
  the physical parameters of the molecular clouds. Their model shows that cooler clouds 
  form preferably lower mass stars. The IMF has a single maximum and an exponential cutoff 
  below it. For the idea of  Favata et al. to be valid, regions with 
  [Fe/H] $<-0.4$ should form stars with an IMF cutoff just below 1 $M_{\sun}$, and in 
  more metal-rich clouds the IMF cutoff should lie beyond 0.6--0.7 $M_{\sun}$. Using 
  the expressions given by Padoan et al. (\cite{padoan}), and taking average values 
  for cloud density and velocity dispersion, the temperature of the clouds 
  for such cutoffs should be 22 K and 19--17 K, respectively. This is hotter than the mean  
  temperature expected for typical dark clouds, 8--15 K (Goldsmith \cite{gold}). However, 
  according to Lin (\cite{lin}), at the present metallicity of the globular clusters ([Fe/H] 
  $\la -1.0$ dex), the cold dense  clouds could cool to around 10 K, putting the IMF cutoff 
  at 0.2 $M_{\sun}$, according to the formulae by Padoan et al. (\cite{padoan}). 

  Even if the IMF cutoff were around 0.9--1 $M_{\sun}$ in the hotter clouds, there would be 
  no such a direct relation between the metallicity and the cloud temperature. The 
  temperature in a molecular cloud is not solely determinated by the cooling rate (which 
  can depend on the metallicity), but it depends also on the cloud density and on the 
  existence of internal and external heating sources (Goldsmith \cite{gold}; Cernicharo 
  \cite{cernicharo}). A difference of 5 K, as that required for the IMF cutoff to be 
  1 $M_{\sun}$ or 0.6 $M_{\sun}$, could exist even inside the same cloud,  where the 
  metallicity is likely to be the same everywhere, as shown by Young et al. 
  (\cite{young}) and Cernicharo 
  (\cite{cernicharo}). There is no strong evidence that the star formation mechanisms would 
  be different for G and K dwarfs. The bump at 0.7 $M_{\sun}$ in the present-day mass 
  function, quoted by Favata et al. (\cite{favata1}) as an evidence favouring a bimodality 
  in the star formation of low mass stars, was more easily explained by Kroupa et al. 
  (\cite{kroupa}) as a real feature in the mass--magnitude relation due to the effects 
  of the increasing importance of ${\rm H}^-$ as an opacity source.
  Given the considerations above, it is reasonable to conclude that MESF cannot account for 
  the lack of metal-poor K dwarfs in the sample by Favata et al. (\cite{favata1}).

  \subsection{A metallicity bias in the catalogue of nearby stars}

  According to Favata et al. (\cite{favata1}), the use of photometric 
  parallaxes could introduce a metallicity bias in the CNS2. Note, however, that our 
  sample does not show this problem, in analogy with the K dwarf metallicity 
  distribution found by Flynn \& Morell (\cite{flynn}). The samples by RPM and 
  Flynn \& Morell were also selected from the Catalogue of Nearby Stars, although 
  both papers have considered a more recent version.
                                                
   \begin{figure}
      \vspace{5cm}
      \caption[]{Sources for the parallaxes in the Third Catalogue of Nearby Star 
       (Gliese \& Jahrei\ss\  \cite{glijar2}).}
         \label{pisources}
   \end{figure}

 \begin{figure}
      \vspace{4.5cm}
      \caption[]{The real inclusion limit of the CNS3 as a function
      of [Fe/H] and $(B-V)$ for {\it UBVRI} parallaxes. The curves 
      correspond to $(B-V)$ of 0.5, 0.6, 0.7, 0.8 and 0.9. The labels 
      indicate the curves for the cooler and hotter stars.}
         \label{25pclim}
   \end{figure}

  We decided to investigate the parallax sources in CNS3. This version 
  of the catalogue was used instead of CNS2, as all recent work 
  on the metallicity distributions is based on it. Moreover, any bias 
  in the CNS2 would also be present in the CNS3, since both 
  catalogues were built in the same fashion. We begin by selecting 
  all stars with $(B-V)$ between 0.5 and 1.4, as in  Favata et al. 
  (\cite{favata2}). The sample was further divided into `G stars' and 
  `K stars' at $(B-V)=0.8$. There are 1421 objects in this colour 
  range, from which 550 are G stars and 871 K stars. Figure 
  \ref{pisources} shows the number of stars included in the CNS3 
  according to the parallax sources. These sources are: (i) 
  trigonometrical parallaxes; (ii) spectroscopic parallaxes and 
  parallaxes determined from broad-band photometric 
  colours; (iii) photometric parallaxes determined from {\it uvby} 
  colours; (iv) photometric parallaxes determined from other photometric 
  systems; and (v) photometric parallaxes for white dwarfs. As can 
  be seen, the main sources for the CNS3 are the 
  trigonometrical parallaxes, and parallaxes determined from spectral 
  types or {\it UBVRI} colours (which we will call {\it UBVRI} 
  parallaxes). The contribution by photometric 
  parallaxes at this colour range is negligible. Both the spectroscopic 
  and {\it UBVRI} parallaxes are determined from mean calibrations 
  built using the stars for which accurate trigonometric parallaxes 
  are available (Gliese \& Jahrei\ss\ \cite{glijahr89}). As these 
  calibrations include stars with varying chemical 
  composition, this must refer to an average metallicity. At a given 
  colour, metal-poor stars have higher absolute magnitudes than their 
  richer counterparts, because their main sequences lay below that 
  of the average-metallicity stars in the colour-magnitude 
  diagram. Therefore, metal-poor stars would be estimated to be 
  systematically farther away than they really are by the use of 
  spectroscopic and {\it UBVRI} parallaxes, as Favata et al. 
  (\cite{favata1}) suggested. Could this effect be large enough 
  to introduce a metallicity bias in the CNS3?

  In order to investigate this problem, we need to know how 
  the `25 pc limit' for inclusion in the 
  CNS3 depends on the metallicity as well on the colour of the stars 
  by using an average colour--magnitude relation.
  We have used the theoretical zero-age main 
  sequences (ZAMS) calculated by VandenBerg (\cite{vandenberg}).
  His ZAMS for ${\rm [Fe/H]} = -0.23$ was chosen as the mean ZAMS, since
  this metallicity corresponds roughly to the average metallicity of 
  the solar neighbourhood stars (cf. RPM). In Figure \ref{25pclim}, 
  we show the real limit for inclusion in the CNS3, 
  for $(B-V)$ colours ranging from 0.50 to 0.90. The figure shows that 
  metal-poor stars, with [Fe/H] $< -0.4$, estimated as being located 
  at 25 pc from the Sun, are in fact closer by 2.5--8 pc. Also, 
  solar-metallicity stars assumed to be within 25 pc from the Sun, 
  could be farther away by up to 5 pc. This effect depends slightly 
  on the stellar colour, being lower for cooler stars. Thus, it is 
  expected that such effects would be slightly more 
  pronounced amongst the G dwarfs, in comparison with the K dwarfs.
     
   \begin{figure*}
      \vspace{10cm}
      \caption[]{Stellar Distances from the CNS3 and from the 
   HIPPARCOS database, for different stellar groups defined by 
   their parallax sources. The dot-dashed lines at the bottom 
   panels separate stars with good distance estimates in the CNS3 from 
   those assumed to be closer than they are.}
         \label{cnship}
   \end{figure*}

  We have looked for such effects in the data by comparing the 
  distances from the CNS3 with the distances measured by the HIPPARCOS 
  satelite, both for  stars with trigonometric and {\it UBVRI} 
  parallaxes. The sample of G and K dwarfs, built according to the
  prescriptions above, was further divided into four samples: (i) 
  G dwarfs included in the CNS3 with trigonometric parallaxes 
  (hereafter tG); (ii) G dwarfs with spectroscopic and {\it UBVRI} 
  parallaxes (ubvG); (iii) K dwarfs included with trigonometric 
  parallaxes (tK); and (iv) K dwarfs with spectroscopic and 
  {\it UBVRI} parallaxes (ubvK). The number  of stars with distances 
  in both the CNS3 and in the HIPPARCOS database is 236 (tG), 204 (ubvG), 
  262 (tK) and 272 (ubvK). 

  Figure \ref{cnship} shows a comparison of the CNS3 and HIPPARCOS 
  distances of these four groups. A number of trends can be seen in 
  these panels. Let us consider first the two groups included in the 
  CNS3 with trigonometric parallaxes, tG and tK. It is 
  possible to see that the agreement between the CNS3 and HIPPARCOS 
  distances improves as we consider stars closer to the Sun, reflecting 
  the better accuracy of ground-based parallax measurements of nearby 
  objects. A very small number of stars 
  was also included in the catalogue in spite of having 
  trigonometric parallaxes smaller than 0.039. There are 
  nearly 10\% of the stars in each group tG and tK that are 
  located much farther away than 25 pc. This is due to 
  errors in the parallax measurements, so that we do not expect 
  any chemical composition differences 
  between those stars and the stars with accurate distances. 
  The situation is different for the groups ubvG and ubvK, 
  whose distances are shown in the bottom panels of Fig. 
  \ref{cnship}. For these groups, the scatter around the line 
  of same distance does not depend on the actual stellar distance. 
  Such scatter is very likely to be produced by 
  the varying chemical composition of these stars. There is a 
  group of stars with underestimated distances in CNS3, both 
  amongst the G and K dwarfs. We separate these stars by a dot-dashed 
  line. The possibility that the inclusion of metal-rich 
  stars in CNS3 with {\it UBVRI} parallaxes has an important 
  effect can be checked by comparing 
  the metallicity of the stars at both sides of the dot-dashed
  lines in the bottom of Fig. \ref{cnship}.

  To estimate the metallicities we used the same procedures 
  described in Section 2. The number of stars with metallicities 
  in each subgroup is: 185 (tG), 176 (ubvG), 111 (tK) and 121 
  (ubvK). The number of stars deviating from the line of same 
  distance is 21 G dwarfs and 20 K dwarfs.

   \begin{figure}
      \vspace{9.5cm}
      \caption[]{Comparison of the metallicity distributions of 
      the stellar groups included in the CNS3 with different 
      parallax sources: a) groups tG e ubvG; b) groups 
      tK and ubvK.
              }
         \label{histogk}
   \end{figure}

  In Figure \ref{histogk}, we show the metallicity distributions of the 
  groups tG, ubvG, tK and ubvK. There is no indication that the 
  metallicity distribution of  G stars is different at the extreme 
  metallicities, regardless of the parallax source. However, 
  the metallicity distribution of the group ubvG 
  has a remarkable single peak at ${\rm 
  [Fe/H]}\sim -0.20$ dex, which is not present in group tG. 
  This peak is also apparent 
  in the metallicity distributions discussed in Section 2, but 
  it is not clear whether it is caused by something related to 
  the colour--magnitude calibration, since it is also 
  present in Twarog's (\cite{twarog}) distribution which uses 
  very different selection criteria. On the other hand, the 
  metallicity distributions of K dwarfs seem to depend 
  strongly on the parallax sources of the CNS3. A Kolmogorov-Smirnov 
  test indicates that both distributions are different 
  at a significance level of 99.99\%. 
  However, the difference occurs in the opposite sense of what 
  we were expecting as  group tK shows much more metal-rich 
  objects than the group ubvK. Also there seems to be more 
  metal-poor stars amongst the ubvK dwarfs. Therefore, 
  these groups do not show any bias derived from {\it UBVRI} 
  parallaxes, although some excess of metal-rich stars
  is apparent in the group of K dwarfs with trigonometric
  parallaxes. 
 
  However, this result is not conclusive, since the metallicity 
  distribution of  group tK is more strongly dependent on the 
  calibration by Olsen (\cite{olsen84}) than group ubvK 
  (the fraction of stars in these groups that 
  have $(b-y) > 0.550$ is 0.55 and 0.37, respectively). 
  It is worth to note  that the metallicity distribution of 
  group ubvK agrees better with the groups 
  of G dwarfs. The hypothesis that the metallicity 
  distribution of group ubvK is the same as those of groups 
  ubvG and tG can only be rejected at a significance level 
  of 0.2246 and 0.2339, respectively. 

  It is then particularly important to see whether there are 
  differences amongst the groups of deviating stars, that
  is, those objects to the right of the dot-dashed lines
  in Fig. \ref{cnship}, and the remaining groups. The metallicites 
  of the deviating G stars range from $-0.5$ to 
  $+0.1$ dex, with an average around $-0.10$ dex. There is no 
  indication that this group has more metal-rich stars compared 
  to the others. The absence of stars with 
  metallicites lower than $-0.5$ dex can well be ascribed to the 
  size of the sample. As an illustration, the KS test gives 
  significance levels of 0.517 and 0.314 for this distribution 
  not to be taken  from the same population of groups ubvG and tG, 
  respectively. The situation is different for K dwarfs. 
  The group of deviating K dwarfs has metallicities ranging 
  from $-1.6$ to $-0.05$ dex, with an average around $-0.65$ dex. 
  This result is very peculiar since it suggests that the stars 
  which have systematically underestimated distances in the CNS3 
  are metal-poor, while we would expect that metal-poor stars 
  would have overestimated distances according 
  to Figure \ref{25pclim}. However, this question cannot be 
  properly answered because the metallicity of the group of
  deviating K dwarfs  also strongly depends on the metallicity 
  calibration for stars cooler than $(b-y)=0.550$. 

  In spite of that, if such bias is likely to be present in the 
  catalogue, it should occur for both G and K dwarfs, being in
  fact stronger for the hotter stars. The non-existence of such 
  bias amongst the G dwarfs, which have even more accurate 
  photometric metallicites, indicates that 
  it does not affect the content of the CNS3. This can happen 
  because the limit for inclusion of objects in the CNS3 due 
  to spectroscopic and {\it UBVRI} parallaxes is 
  more flexible than the limit for trigonometric 
  parallaxes. This is evident from Figure \ref{cnship}. In this plot 
  we see that there are many stars in the CNS3 whose distances 
  in this catalogue  are greater than 25 pc, 
  amongst those included with {\it UBVRI} parallaxes. Thus, in 
  the CNS3 there is not a fixed limit at 25 pc for the inclusion 
  of stars with {\it UBVRI} parallaxes, and there seems to be no 
  corresponding metallicity-bias.
  
  The simplest hypothesis to account for the results found 
  by Favata et al. (\cite{favata1}) is that their sample is not representative of the
  galactic population of K dwarfs, due to its small size. The original sample randomly selected 
  from the CNS2, and consisting of around 100 G and 100 K dwarfs (Favata et al. 
  \cite{favata2}), can be expected to be representative. However, the number of stars 
  that were effectively observed is 63 G dwarfs and 26 K dwarfs. As Favata et al. 
  (\cite{favata1}) themselves state, relatively fewer cooler stars were observed due 
  to their faint magnitudes. This observational selection is not likely to remove the 
  representativeness of a large data sample. However, small samples are much easily 
  affected by statistical fluctuations due to the elimination of some stars. This can 
  explain why the distributions by Favata et al. (\cite{favata1}) show large 
  fluctuations and not a single prominent peak as the other metallicity distributions in 
  the literature.

\begin{acknowledgements}
      HJR-P thanks Mar\'{\i}lia Sartori for some helpful discussions. 
      We are indebted to Dr. E. H. Olsen for some important remarks
      on an earlier version of this paper. We have made use of the 
      SIMBAD database, operated at CDS, Strasbourg, France. This work 
      has been partially supported by FAPESP and CNPq. 
\end{acknowledgements}

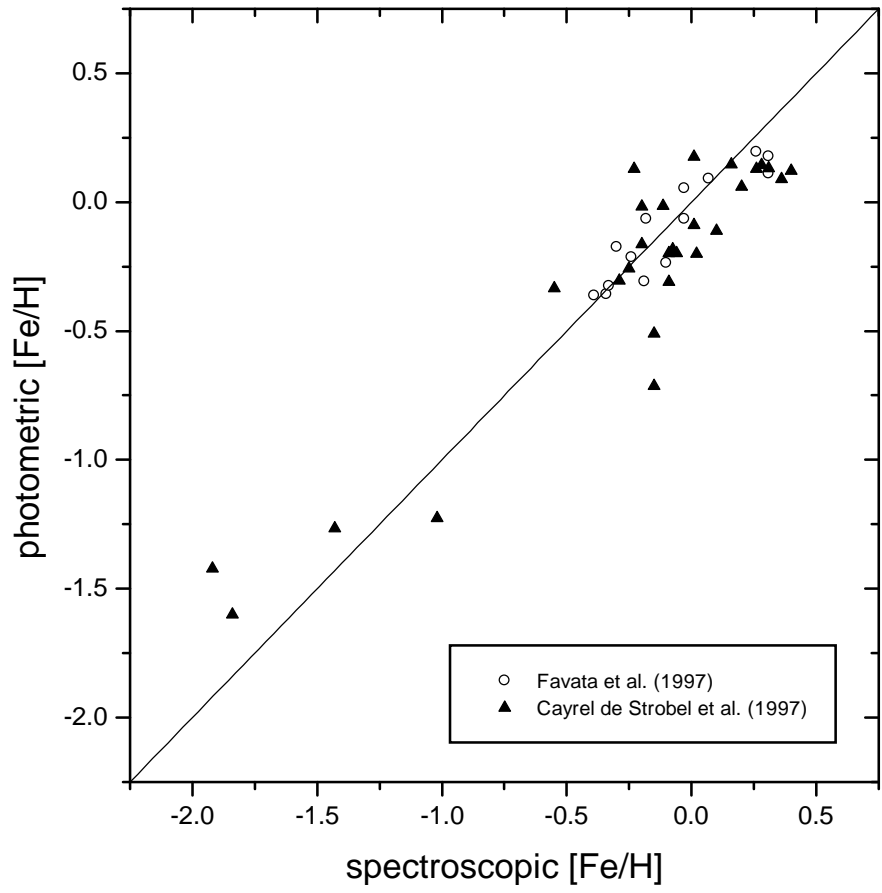
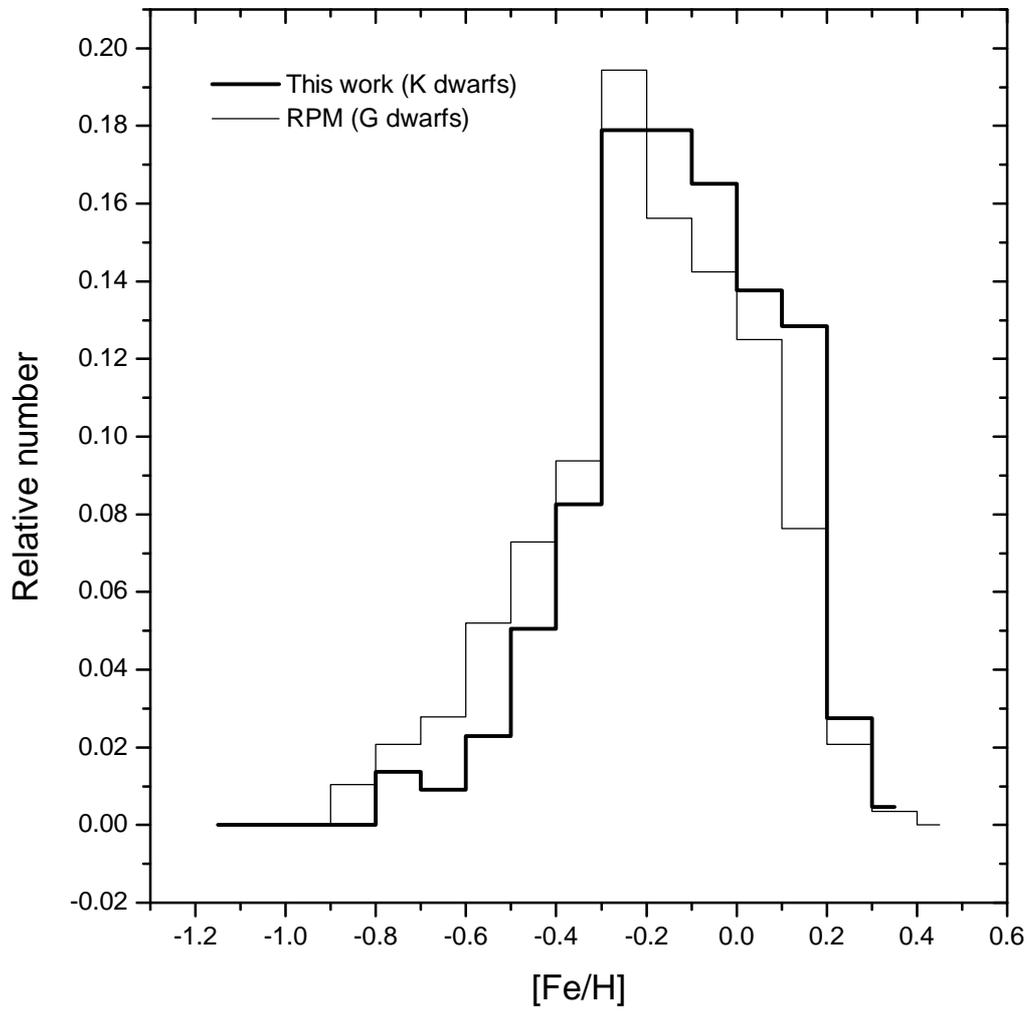
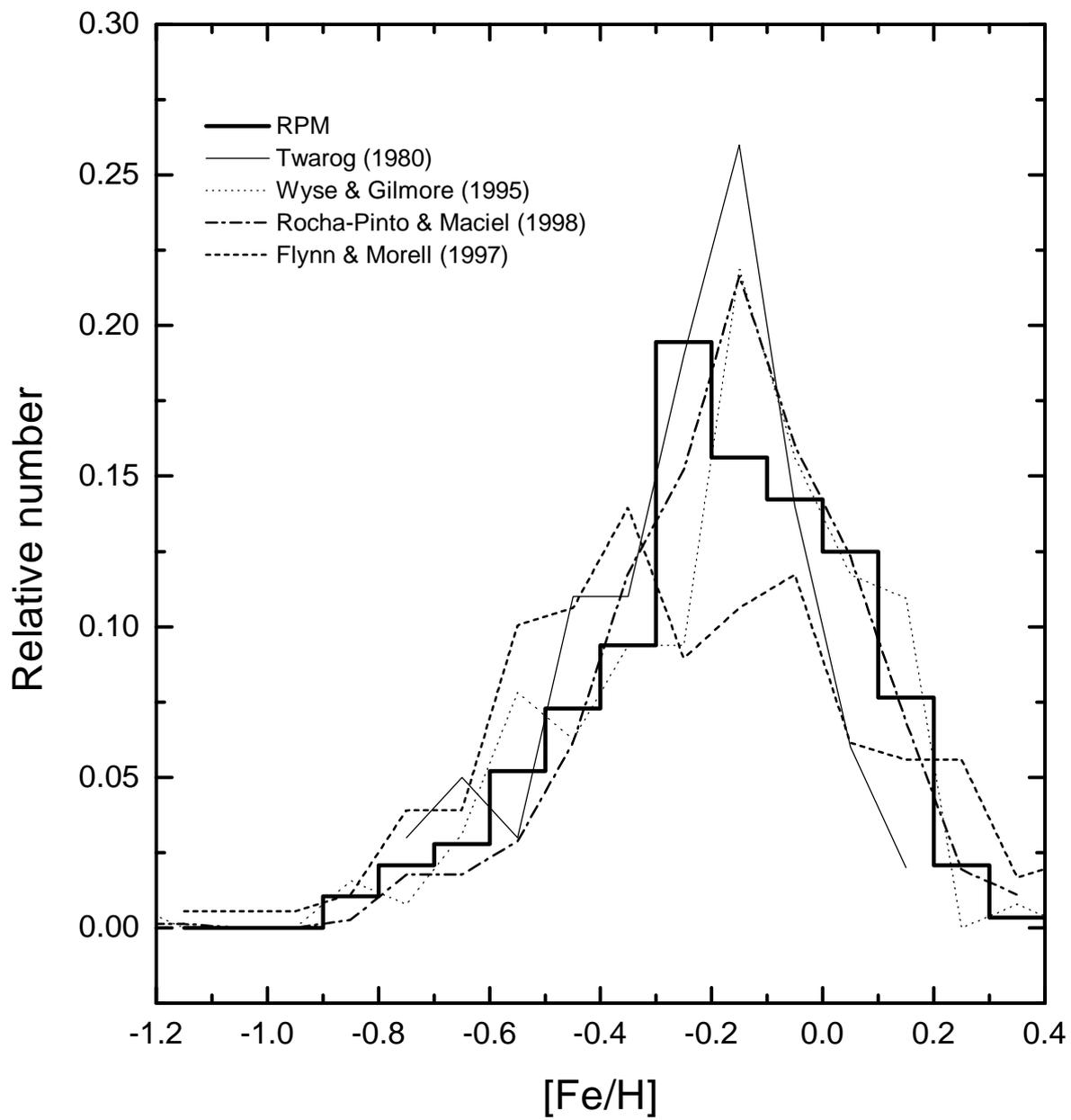
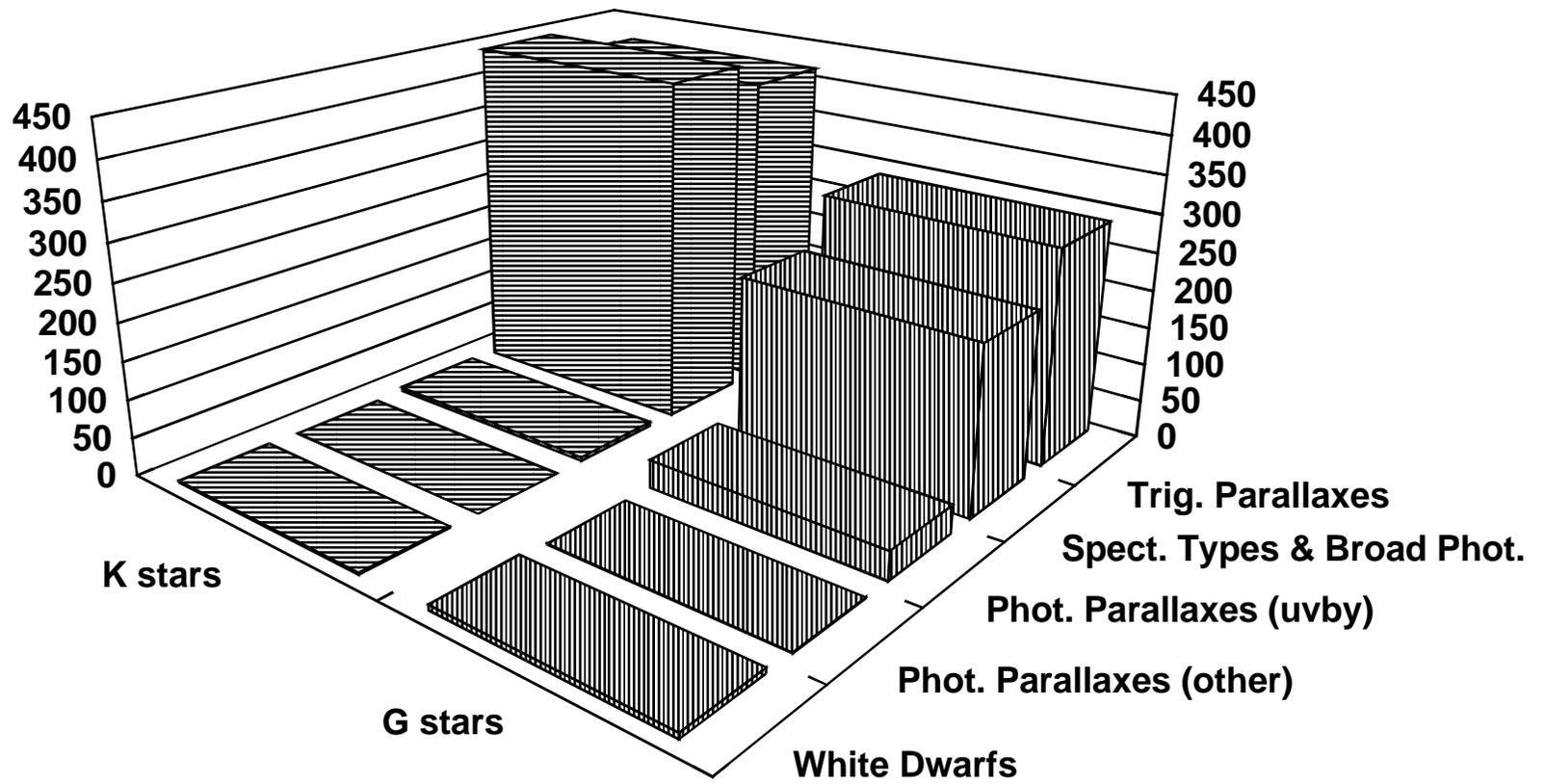
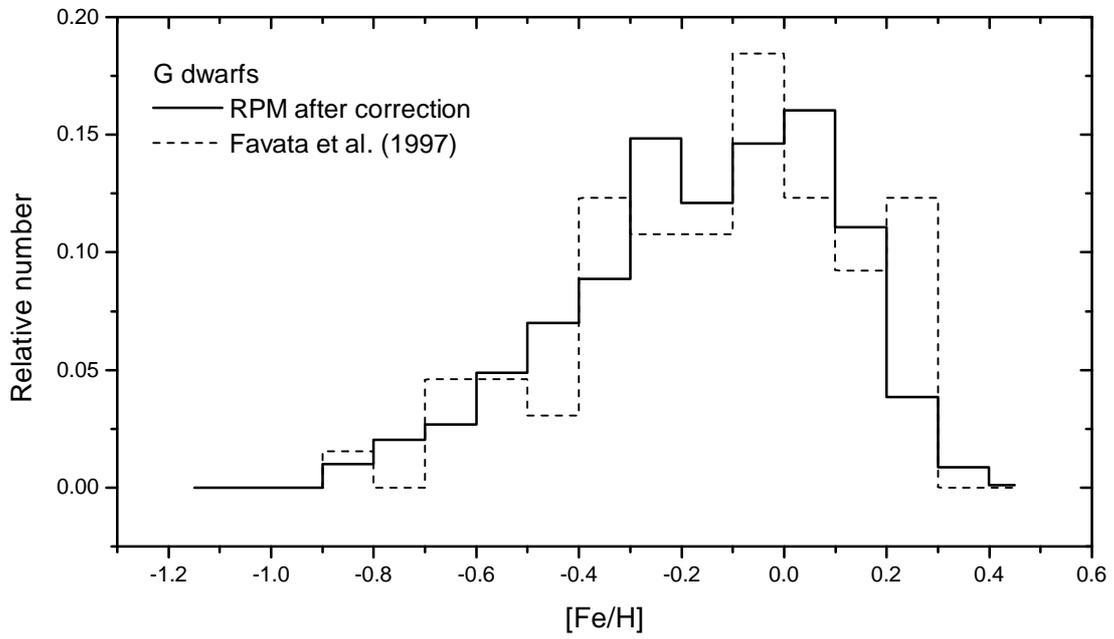
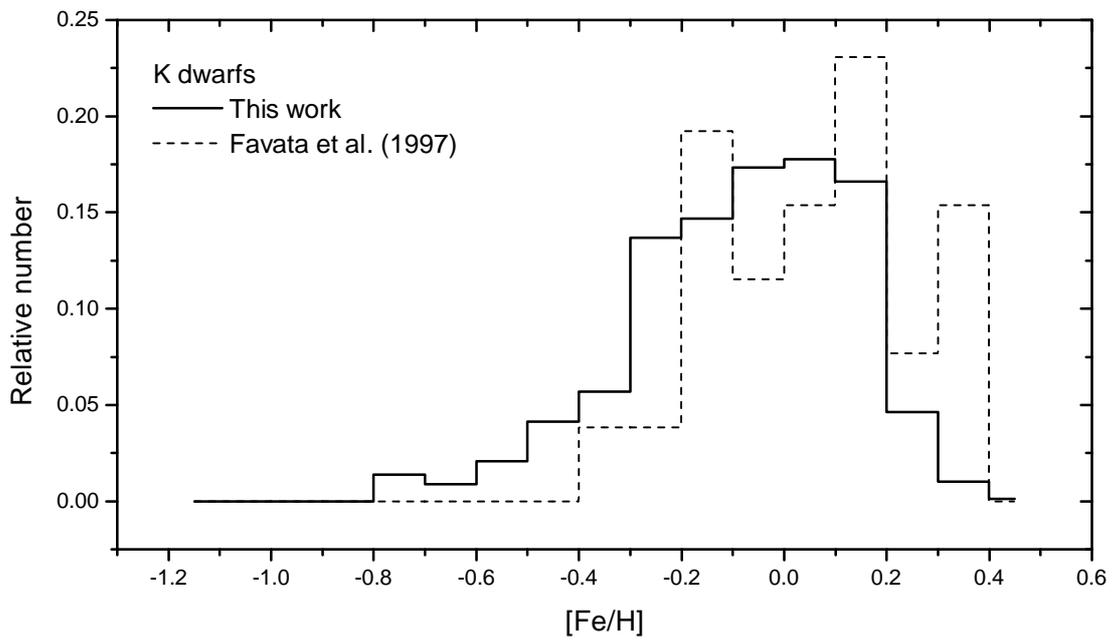
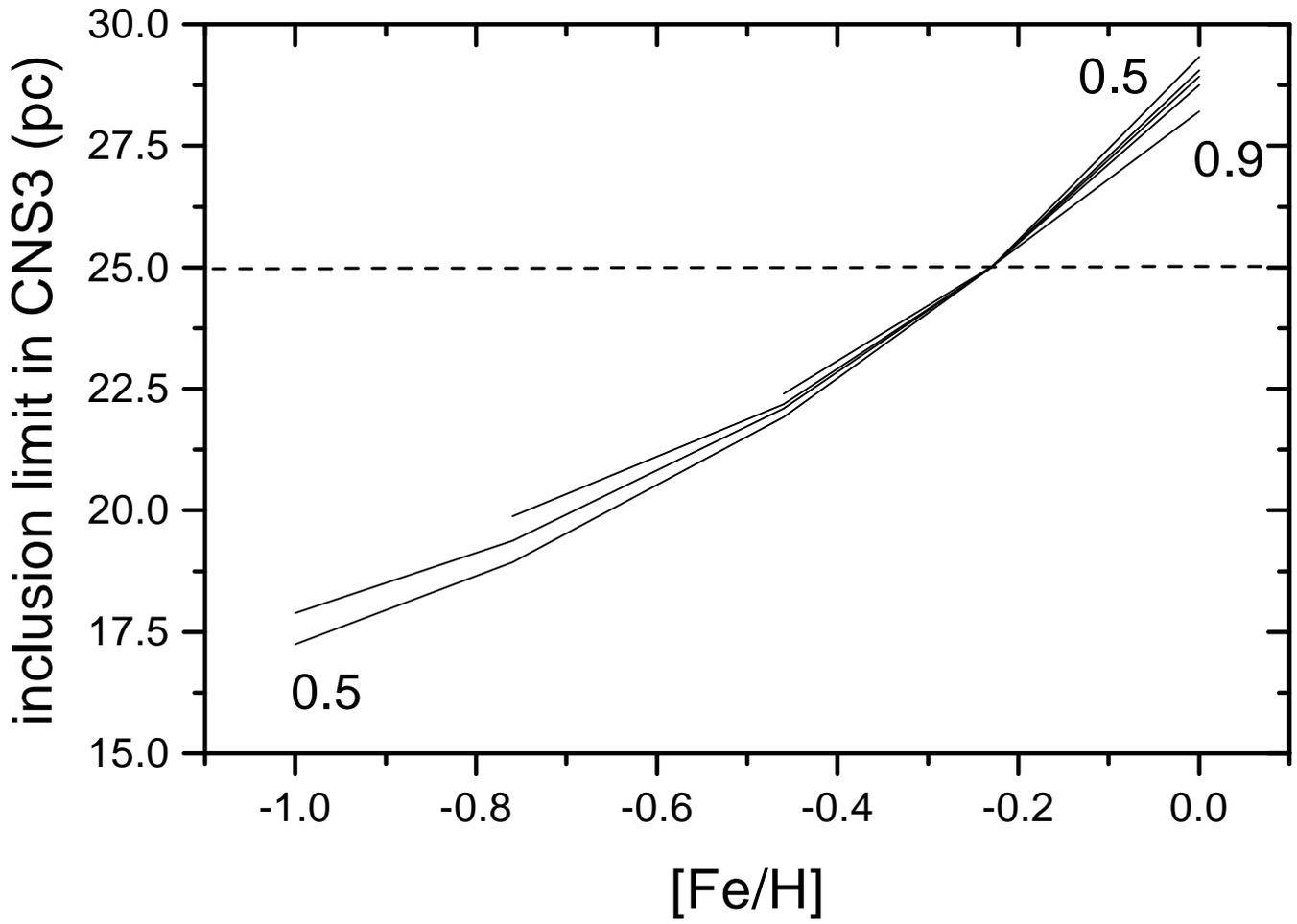
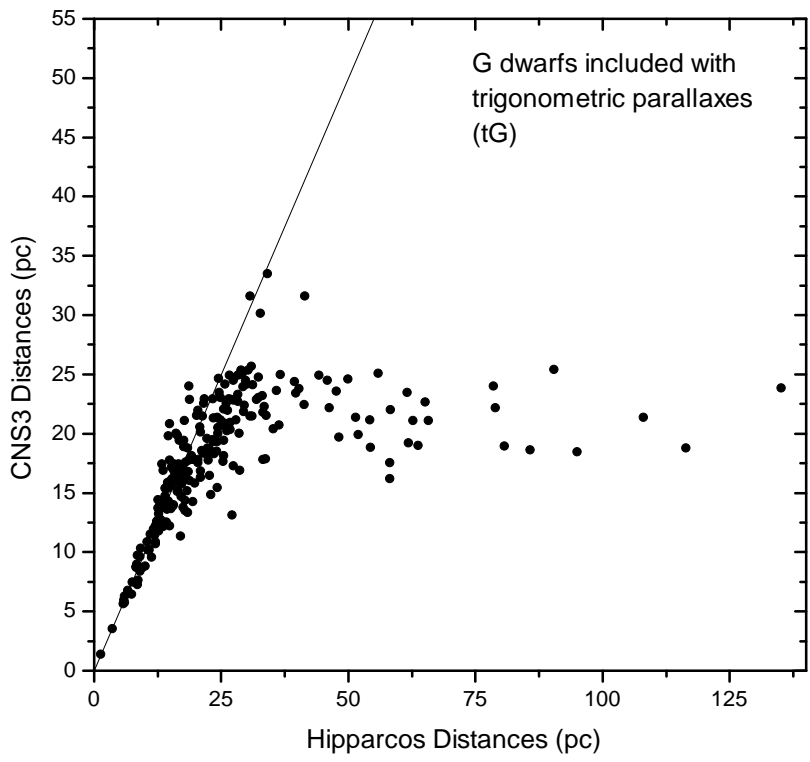
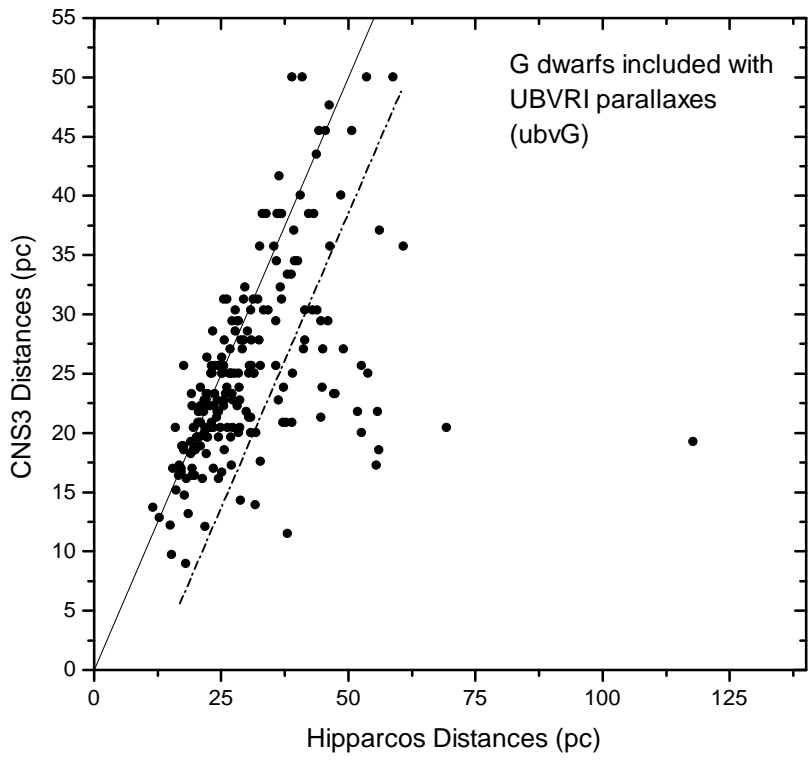
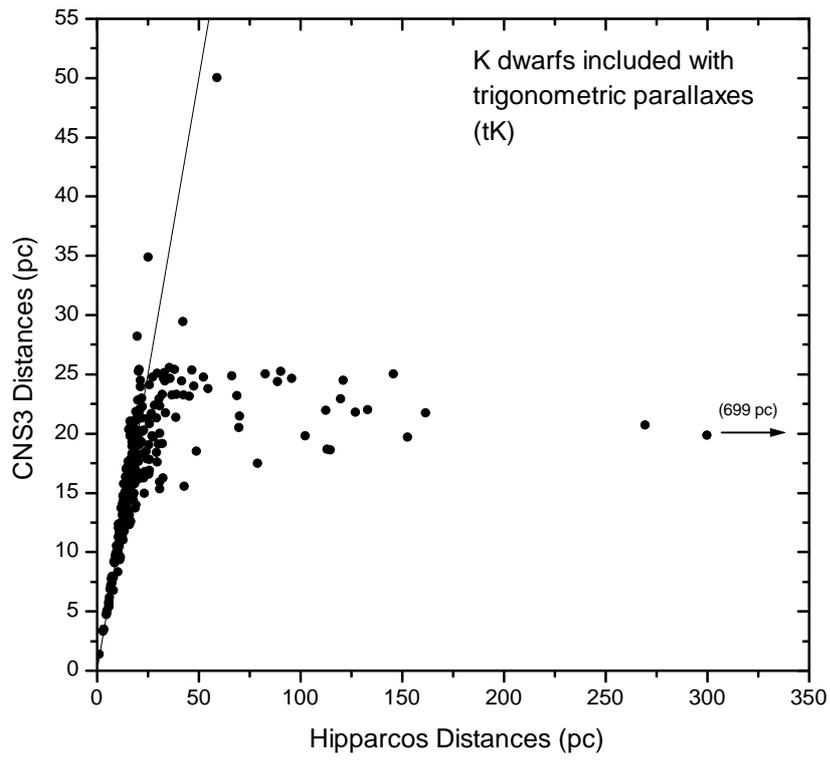
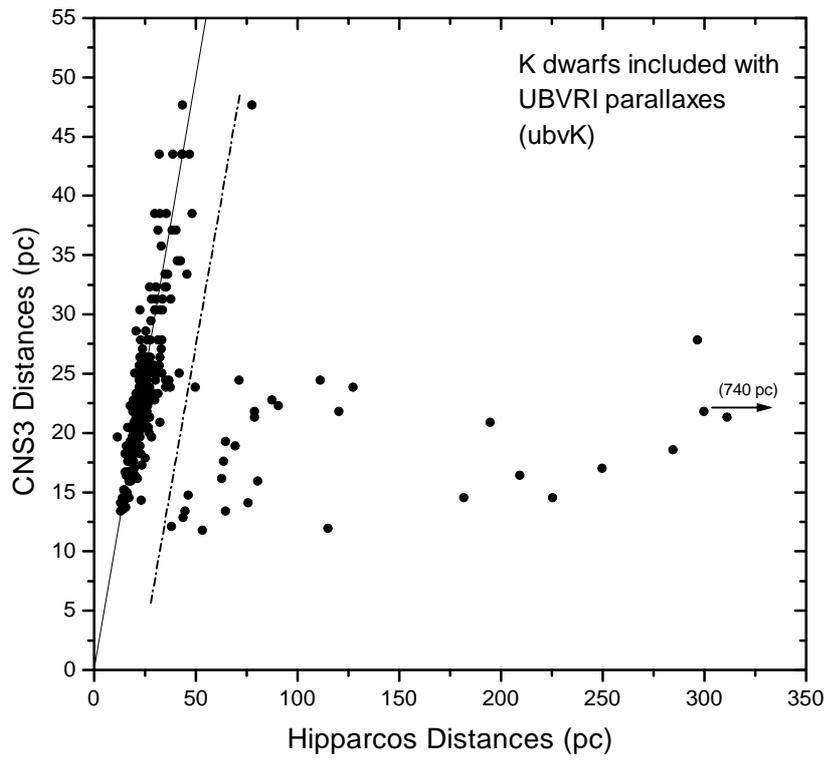
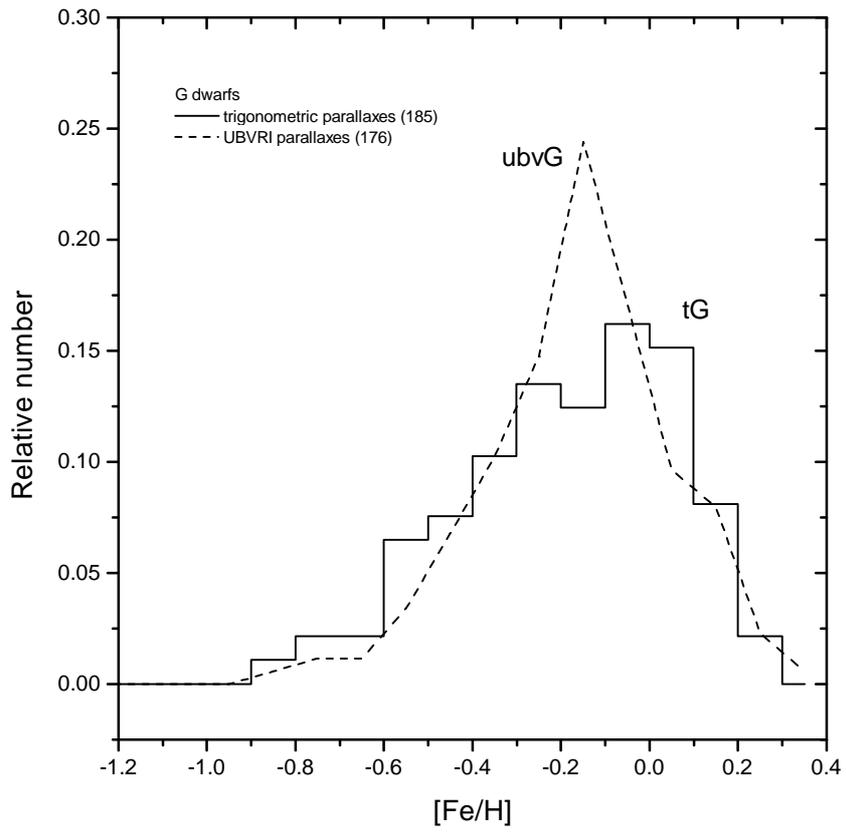
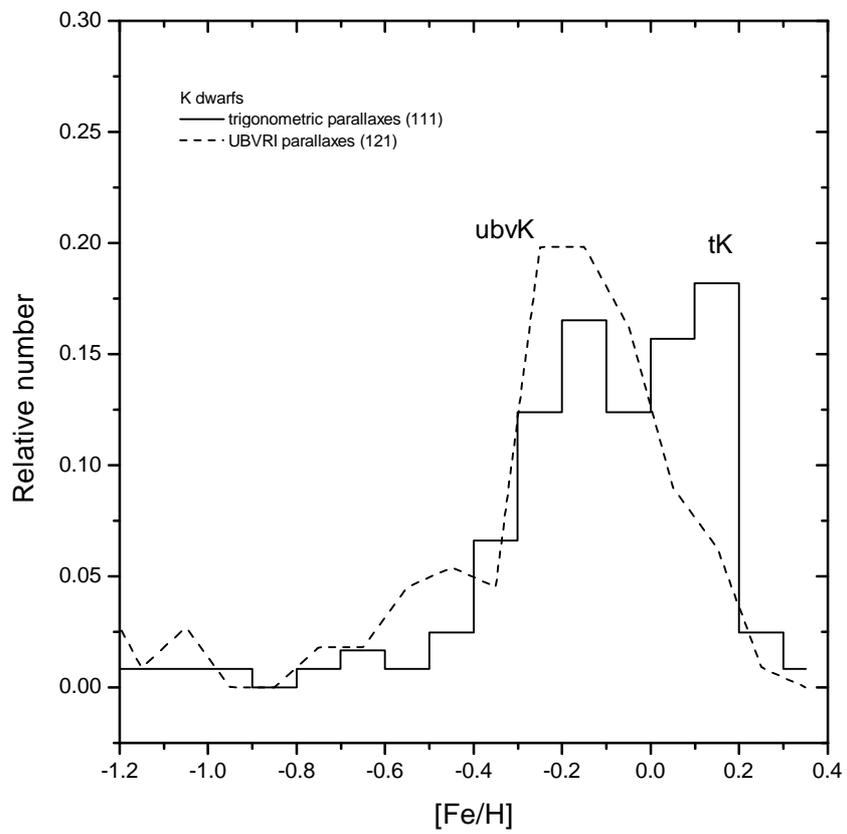
\end{document}